\documentclass{emulateapj}
\usepackage{color}
\usepackage{url}
\usepackage{rotating}
\usepackage{subfigure}

\shorttitle{Bright Merger-Nova Emission from a New-born Black Hole}
\shortauthors{Ma et al.}
\slugcomment{}

\begin{document}

\title{Bright ``MERGER-NOVA'' Emission Powered by Magnetic wind from a New-born Black Hole}
\author{Shuai-Bing Ma$^{1}$, Wei-Hua Lei$^{1}$, He Gao$^2$, Wei Xie$^{3}$, Wei Chen$^{1}$, Bing Zhang$^{4}$, and Ding-Xiong Wang$^{1}$}
\affil{
$^{1}$School of Physics, Huazhong University of Science and Technology, Wuhan 430074, China. Email: leiwh@hust.edu.cn \\
$^{2}$Department of Astronomy, Beijing Normal University, Beijing 100875, China. Email: gaohe@bnu.edu.cn \\
$^{3}$Guizhou Provincial Key Laboratory of Radio Astronomy and Data Processing, Guizhou Normal University, Guiyang, 550001, China. \\
$^{4}$Department of Physics and Astronomy, University of Nevada Las Vegas, NV 89154, USA.
}

\begin{abstract}
Mergers of neutron star-neutron star (NS-NS) or neutron star-black hole (NS-BH) binaries are candidate sources of gravitational waves (GWs). At least a fraction of the merger remnant should be a stellar mass BH with a sub-relativistic ejecta. A collimated jet is launched via Blandford-Znajek mechanism from the central BH to trigger a short gamma-ray burst (SGRB). At the same time, a near-isotropic wind may be driven by the Blandford-Payne mechanism (BP). In previous work, additional energy injection to the ejecta from the BP mechanism was ignored, and radioactive decay has long been thought as the main source of the kilonova energy. In this Letter, we propose that the wind driven by the BP mechanism from the new-born BH-disk can heat up and push the ejecta during the prompt emission phase or even at late time when there is fallback accretion. Such a BP-powered merger-nova could be bright in the optical band even for a low-luminosity SGRB. The detection of a GW event with a merger product of BH, and accompanied by a bright merger-nova, would be a robust test of our model.

\end{abstract}

\keywords{gamma-ray burst: general---gravitational waves}

\section{Introduction}
The detection of gravitational waves (GWs) from mergers of binary black holes (BHs) by Advanced LIGO has started the era of GW astronomy (Abbot et al. 2016a,b). Searching for electromagnetic (EM) counterparts to GW events is of great interest, since it could play a crucial role in locating the host galaxy and studying detailed physics of compact star mergers (Fan et al. 2017). Systematic searches for EM signals has been performed to each GW event. However, general consensus for the production mechanism of EM counterparts to BH-BH mergers is still lacking. Very recently, advanced LIGO and VIRGO detected GW170817, which is consistent with an NS-NS merger. Subsequently, the multi-wavelength EM counterparts (GRB 170817A, AT2017gfo) were identified. This marked the beginning of the era of multi-messenger astronomy (Abbot et al. 2017a,b). 

Short-duration gamma-ray bursts (SGRBs) are probably the brightest EM emission during the NS-NS or NS-BH mergers (Gehrels et al. 2005; Fox et al. 2005; Fong et al. 2013; Virgili et al. 2011). The discovery of the association of GW170817 with GRB 170817A directly confirmed this scenario (Abbot et al. 2017b). However, due to the collimation of GRB jets, not all NS-NS GW sources will be associated with SGRBs (e.g., Burrows et al. 2006). The ``kilonova'' powered by radioactive decay of r-process is then of great importance (Li \& Paczy\'nski 1998; Metzger et al. 2010), which has been broadly adopted to interpret AT2017agfo, the optical counterpart of GW170817 (e.g. Kasen et al. 2017).

If the merger product is a millisecond magnetar, the near-isotropic magnetar wind would produce rich EM signals in association with a GW event. Zhang (2013) proposed that such a magnetar wind would undergo magnetic dissipation (Zhang \& Yan 2011) and power a bright X-ray afterglow emission. In addition, a significant fraction of the magnetar wind energy would be used to heat up and push the neutron-rich ejecta, which powers a bright ``merger-nova'' (could be much brighter than ``kilonova'', Yu et al. 2013; Metzger \& Piro 2014; Gao et al. 2015, 2017) and broad-band afterglow (in radio, optical and X-ray, see Gao et al. 2013).

However, the merger product of BH-NS mergers and at least a fraction of NS-NS mergers is a stellar mass BH (Jin et al. 2015, 2016; Gao et al. 2016). Radioactive decay has long been thought as the main source of energy for ``kilonova'' in the BH scenario. In this work, we propose that the magnetic wind from the accretion disk would heat up the neutron-rich merger ejecta and produce a bright ``merger-nova''. We study the emission properties of such a magnetic-wind-powered merger-nova for a new-born BH with and without a fall-back accretion disk at late time. The predicted EM signals can serve as interesting targets in the search for EM counterparts of GW burst triggers in the Advanced LIGO/Virgo era.

\section{Magnetic Wind from BH central engine}
The mergers of NS-BH binaries and at least a fraction of NS-NS binaries would lead to the formation of a hyper-accreting stellar mass BH with ejecta of mass $M_{\rm ej} \sim 10^{-3} - 10^{-1} M_\sun$. The GRB prompt emission can be powered by the Blandford \& Znajek (1977, hereafter BZ) mechanism, in which the spin energy of the BH is extracted via the open field lines penetrating the event horizon. For a BH with mass $M_\bullet$, spin $a_\bullet$ and accretion rate $\dot{M}$, the BZ power can be estimated as (Lei et al. 2013; Lei et al. 2017; Liu et al. 2017),
\begin{equation}
L_{\rm BZ}=1.7 \times 10^{50}  a_{\bullet}^2 m_{\bullet}^2
B_{\bullet,15}^2 F(a_{\bullet}) \ {\rm erg \ s^{-1}},
\end{equation}
where $m_\bullet = M_\bullet/M_\sun$ and $F(a_\bullet) = [(1+q^2)/q^2] [(q+1/q)\arctan q-1]$, and $q=a_\bullet/(1+\sqrt{1-a_\bullet^2})$.

As the magnetic field on the BH is supported by the surrounding disk, there are some relations between $B_\bullet$ and $\dot{M}$. As a matter of fact, these relations might be rather complicated, and would be very different in different situations. It is reasonable to assume that the magnetic pressure on the horizon may reach a fraction $\alpha_{\rm m}$ of the ram pressure of the innermost parts of an accretion flow, i.e.,
\begin{equation}
B_\bullet^2 /8\pi = \alpha_{\rm m} P_\mathrm{ram} \sim \alpha_{\rm m} \dot{M} c/(4\pi r_\bullet^2).
\label{eq7}
\end{equation}

The observed $\gamma$-ray/X-ray luminosity is connected to the BZ power via the X-ray radiation efficiency $\eta$ and the jet beaming factor $f_{\rm b}$, i.e.,
\begin{equation}
\eta L_{\rm BZ}=f_{\rm b} L_{\rm \gamma,iso}
\label{eq2}
\end{equation}

Energy and angular momentum could also be extracted magnetically from the accretion disk, by field lines that leave the disk surface and extend to large distances, centrifugally launching a baryon-rich wide wind/outflow through the Blandford-Payne (Blandford \& Payne 1982, hereafter BP)
mechanism. The magnetic wind power can be estimated by (Livio et al. 1999; Meier 2001)
\begin{equation}
L_\mathrm{BP}=(B_\mathrm{ms}^\mathrm{p})^2r_\mathrm{ms}^4\Omega_\mathrm{ms}^2/32c
\label{eq:Lbp}
\end{equation}
\noindent where $\Omega_\mathrm{ms}$ is the Keplerian angular velocity at the radius of the marginally stable orbit $r_\mathrm{ms}$.
The expression for $r_{\rm ms}$ is from (Bardeen et al. 1972),
\begin{eqnarray}
r_{\rm ms}/r_{\rm g} =  3+Z_2 -\left[(3-Z_1)(3+Z_1+2Z_2)\right]^{1/2},
\label{eq4}
\end{eqnarray}
for $0\leq a_{\bullet} \leq 1$, where $Z_1 \equiv 1+(1-a_{\bullet}^2)^{1/3} [(1+a_{\bullet})^{1/3}+(1-a_{\bullet})^{1/3}]$, $Z_2\equiv (3a_{\bullet}^2+Z_1^2)^{1/2}$. The Keplerian angular velocity is given by
\begin{equation}
\Omega_\mathrm{ms}=\left(\frac{GM_\bullet}{c^3}\right)^{-1}\frac{1}{\chi^{3}_\mathrm{ms}+a_\bullet},
\label{eq5}
\end{equation}
where $\chi_\mathrm{ms}\equiv\sqrt{r_\mathrm{ms}/r_\mathrm{g}}$, and $r_\mathrm{g}\equiv G M_\bullet/c^2$. Following Blandford \& Payne (1982), the disk poloidal magnetic field $B_\mathrm{ms}^\mathrm{p}$ at $r_\mathrm{ms}$ can be expressed as,

\begin{equation}
B_\mathrm{ms}^{\rm{p}} = B_\bullet (r_\mathrm{ms}/r_\bullet)^{-5/4}
\label{eq6}
\end{equation}
\noindent where $r$ is the disk radius and $r_\bullet = r_{\rm g} (1+\sqrt{1-a_\bullet^2})$ is BH horizon radius. $B_\bullet$ is the magnetic field strength threading the BH horizon.

Since the BH would be spun up by accretion and spun down by the BZ mechanism during a GRB, the evolution equations of a Kerr BH can be written as,
\begin{equation}
\frac{dM_\bullet}{dt} = \dot{M} E_\mathrm{ms}^\dag - L_{\mathrm{BZ}},
\label{eq12}
\end{equation}

\begin{equation}
\frac{dJ_\bullet}{dt} = \dot{M} J_\mathrm{ms}^\dag - L_{\mathrm{BZ}}/(0.5 \Omega_{\bullet}),
\label{13}
\end{equation}
where $J_\bullet=a_\bullet G M_\bullet^2/c$ is the angular momentum of BH, and $\Omega_\bullet =a_\bullet c/(2 r_\bullet)$ is the angular velocity of the BH horizon. $E_\mathrm{ms}^\dag=(4\chi_\mathrm{ms}-3a_\bullet)/\sqrt{3}\chi_\mathrm{ms}^2$ and $J_\mathrm{ms}^\dag=2GM_\bullet (3\chi_\mathrm{ms}-2a_\bullet)/\sqrt{3}c\chi_\mathrm{ms}$ are the specific energy and specific angular momentum of a particle at $r_{\rm ms}$, respectively (Novikov \& Throne 1973).

We use a simple model to describe the evolution of the accretion rate during the prompt emission phase (see also Kumar et al. 2008), in which the disk with mass $M_{\rm d}$ and angular momentum $J_{\rm d}$ is treated as a single annulus ring with effective disk radius $r_{\rm d}=J_{\rm d}^2/(G M_\bullet M_{\rm d}^2)$. The typical accretion time-scale is $t_{\rm acc}=r_{\rm d}^2 / \nu \sim 2/(\alpha \Omega_K)$, where $\alpha$ is the dimensionless viscosity parameter. We thus define the accretion rate as
\begin{equation}
\dot{M} = M_{\rm d}/t_{\rm acc}.
\label{Eq:dMacc}
\end{equation}
The BP driven wind will take away significant angular momentum from the disk, while its associated mass loss rate can be neglected (Li et al. 2008). The disk therefore evolves with time $t$ as
\begin{equation}
\frac{dM_\mathrm{d}}{dt} = -\dot{M},
\label{eq9}
\end{equation}

\begin{equation}
\frac{dJ_\mathrm{d}}{dt} = -\dot{M} J_\mathrm{ms}^\dag-L_\mathrm{BP}/\Omega_\mathrm{ms},
\label{eq10}
\end{equation}

\begin{figure}[htp]
\centering
\includegraphics[width=8cm,angle=0]{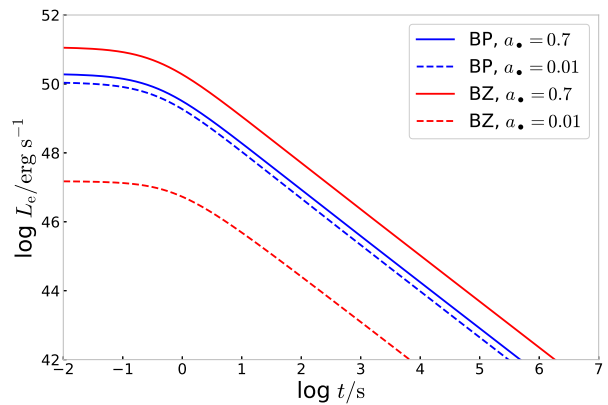}
\caption{Time evolutions of the wind power $L_\mathrm{BP}$(blue lines) for the BHs with $a_{\bullet}=0.7$ (solid lines) and $a_{\bullet}=0.01$ (dashed lines). For comparison, we also show the results of BZ power $L_\mathrm{BZ}$(red lines). In the calculations, the initial BH mass $M_{\bullet}(0)=3M_\odot$, initial disk mass $M_{\mathrm d}(0)=0.01M_{\odot}$, viscosity parameter $\alpha=0.1$ and the parameter $\alpha_{\rm m}=1$ are adopted.}
\label{fig1}
\end{figure}

In Figure \ref{fig1}, we plot the time dependent BP wind (blue lines) power during the prompt emission phase for high ($a_\bullet=0.7$, solid lines) and low initial BH spin ($a_\bullet=0.01$, dashed lines) cases, and compare it with the BZ jet power (red lines). As we can see, the BP wind power is insensitive to the BH spin and dominates over the BZ jet power for BHs with a very small spin. This may account for the relative bright mergernova but weak $\gamma$-ray power for GRB 170817A.

The observations of flares, plateaus and giant bumps suggest that some GRB central engines are long-lived (Burrows et al. 2005; Wu et al. 2013; L\"u et al. 2015; Chen et al. 2017). In the context of BH central engine, fall-back accretion of the surrounding unbounded matter can be a natural interpretation for these activities in SGRBs.

We also study the case with fall-back accretion. We assume that the disk accretion at late time tracks the fall-back accretion rate (MacFadyen et al. 2001; Dai \& Liu 2012),
\begin{eqnarray}
\dot{M} = \dot{M}_{\rm p} \left[ \frac{1}{2}\left(\frac{t-t_0}{t_{\rm p}-t_0} \right)^{-1/2} +  \frac{1}{2}\left(\frac{t-t_0}{t_{\rm p}-t_0} \right)^{5/3} \right]^{-1},
\label{dotm14}
\end{eqnarray}
where $t_0$ is the beginning time of the fall-back accretion, $t_{\rm p}$ is the time corresponding to the peak fall-back rate $\dot{M}_{\rm p}$. As an example, we take $t_0=1000$s, $t_{\rm p}=1500$s(moderate values for X-ray flares in SGRBs, e.g., GRB 050724 and GRB 110731A, see L\"u et al. 2015 and Chen et al. 2017) and $\dot{M}_{\rm p}=6 \times 10^{-6} M_{\sun}~s^{-1}$ in the following calculations. 

\section{Merger-nova Emission from ejecta with energy injection from the BP wind}
The near isotropic magnetic wind dissipates a fraction of the energy into the merger ejecta (Bucciantini et al. 2012). The deposited energy would increase the internal energy and accelerate the ejecta (Yu et al. 2013).
The total energy of the ejecta is thus written as
\begin{equation}
E_{\rm ej}=(\Gamma-1)M_{\rm ej}c^2+\Gamma E'_{\rm int}
 \label{eqs:Eej}
\end{equation}
where $\Gamma$ is the Lorentz factor and $E_{\rm int}^\prime$ is the internal energy measured in the comoving rest frame. The conservation of energy conservation reads,

\begin{equation}
dE_{\rm ej}=( L_{\rm inj}-L'_{e})dt,
 \label{eqs:dEej}
\end{equation}
$ L_{e} $ is bolometric radiation luminosity from the heated electrons. $L_{\rm inj}=\xi L'_{\rm BP}+L'_{\rm ra}$ denotes the injected power from a BP wind $\xi L'_{\rm BP}$ and from an radioactive decay rate $L'_{\rm ra}$. Here we introduce a parameter $\xi$ to describe the fraction of magnetic wind energy that is used to heat the ejecta. The case with $\xi=0$ or $\alpha_{\rm m}=0$ will return to the result of kilonova. Following Yu et al. (2013), we take a normal value $\xi=0.3$.

Combining Equations (\ref{eqs:Eej}) and (\ref{eqs:dEej}), we obtain the dynamic evolution of the ejecta,
\begin{equation}
\frac{d\Gamma}{dt}=\frac{ L_{\rm inj}-L_{e}-\Gamma{\cal D}(dE'_{\rm int}/dt')}{M_{\rm ej}c^2+E'_{\rm int}}.
\end{equation}
where $ {\cal D}=1/[\Gamma(1-\beta)] $ is the Doppler factor, and $ \beta=\sqrt{1-\Gamma^{-2}}$.
The change of the internal energy can be written as (Kasen \& Bildsten 2010)

\begin{equation}
\frac{dE'_{\rm int}}{dt'}= L'_{\rm inj}-L'_e-P'\frac{dV'}{dt'}
\end{equation}

The co-moving luminosities are defined as $ L'_{\rm inj}=L_{\rm inj}/{\cal D}^2 $, and
\begin{eqnarray}
\lefteqn{L'_{\rm ra}=4\times 10^{49}M_{\rm ej,-2}}
\nonumber\\
& & \times\left[\frac{1}{2}-\frac{1}{\pi}\arctan\left(\frac{t'-t'_0}{t'_{\sigma}} \right) \right]^{1.3}{\rm erg~s^{-1}}.
\end{eqnarray}
with $t'_{0}\sim 1.3$ $\rm s$ and $t'_{\sigma}\sim 0.11$ (Korobkin et al. 2012). For a relativistic gas, the pressure is (1/3) of the internal energy density, i.e., $P'=E'_{\rm int}/(3V')$.

The co-moving volume is determined by $dV'/dt'=4\pi R^2\beta c$ and $dR/dt=\beta c/(1-\beta)$. The co-moving frame bolometric emission luminosity of the heated electrons can be estimated as
\begin{equation}
L'_e=\left \{ \begin{array}{ll}
E'_{\rm int}c/(\tau R/\Gamma), & \textrm{ $t < t_{\tau}$},\\
\\
E'_{\rm int}c/(R/\Gamma), & \textrm{ $t \geq t_{\tau}$},\\
\end{array} \right.
\end{equation}
where $\tau=\kappa (M_{\rm ej}/V')(R/\Gamma)$ is optical depth and $\kappa$ is opacity of the ejecta. $t_{\tau}$ is the time when $\tau=1$.

The peak energy of the emission spectrum $\nu L_{\nu}$ is
\begin{equation}
\varepsilon_{\gamma,p}\approx 4{\cal D}kT'=\left \{ \begin{array}{ll}
4{\cal D}k\left( \frac{E'_{\rm int}}{aV'\tau}\right) ^{1/4} & \textrm{for $\tau>1$}\\
\\
4{\cal D}k\left( \frac{E'_{\rm int}}{aV'}\right) ^{1/4}  & \textrm{for $\tau \leq 1$}\\
\end{array} \right.
\end{equation}
where $k$ is the Boltzmann constant and $a$ is the blackbody radiation constant. The luminosity at a particular frequency $\nu$ is given as
\begin{eqnarray}
\nu L_{\nu}={8\pi^2  {\cal D}^2R^2\over
h^3c^2}{(h\nu/{\cal D})^4\over \exp(h\nu/{\cal D}kT')-1},
\end{eqnarray}

\begin{figure}[ht]
\center
\includegraphics[width=8cm,angle=0]{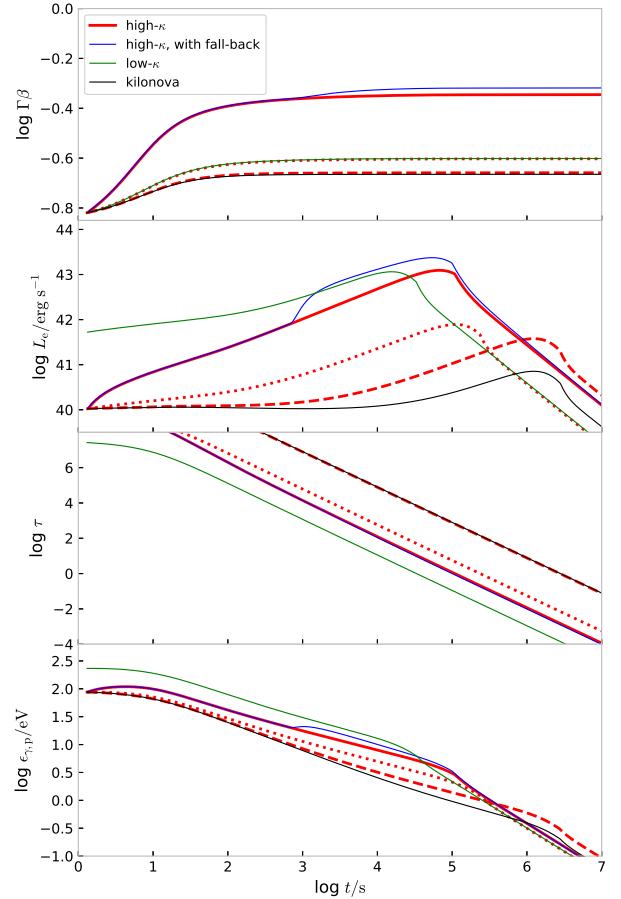}
\caption{The properties of BP-powered merger-nova for different $M_{\rm d}$, $M_{\rm ej}$, $\alpha_{\rm m}$ and $\kappa$: red solid lines ($M_{\rm d}=0.1M_\sun$, $M_{\rm ej}=10^{-3}M_\sun$, $\alpha_{\rm m}=1$, $\kappa=10\rm cm^2 g^{-1}$); red dashed lines ($M_{\rm d}=0.1M_\sun$, $M_{\rm ej}=0.1M_\sun$, $\alpha_{\rm m}=1$, $\kappa=10\rm cm^2 g^{-1}$); red dotted lines ($M_{\rm d}=0.01M_\sun$, $M_{\rm ej}=10^{-3}M_\sun$, $\alpha_{\rm m}=1$, $\kappa=10\rm cm^2 g^{-1}$); green solid lines ($M_{\rm d}=0.1M_\sun$, $M_{\rm ej}=10^{-3}M_\sun$, $\alpha_{\rm m}=0.1$, $\kappa=0.2\rm cm^2 g^{-1}$). The results for a BH with fall-back accretion are shown with blue solid lines ($M_{\rm d}=0.1M_\sun$, $M_{\rm ej}=10^{-3}M_\sun$, $\alpha_{\rm m}=1$, $\kappa=10\rm cm^2 g^{-1}$). The black solid lines ($M_{\rm ej}=0.1M_\sun$, $\alpha_{\rm m}=0$, $\kappa=10\rm cm^2 g^{-1}$) represent the case of a kilonova.}

\label{fig2}
\end{figure}

The properties of such a BP-powered merger-nova are exhibited in Figure \ref{fig2}. For comparison, we present a result with $\alpha_{\rm m}=0$ or kilonova. One can find that the new-born BH can also produce a bright merger-nova once a strong BP-wind is developed from the disk.

A constant opacity $\kappa=0.2 \rm cm^2 g^{-1}$ is adopted by Yu et al. (2013). Kasen et al. (2013), on the other hand, pointed out that heavier elements (particularly the lanthanides) might increase the opacity by several orders of magnitude $\kappa \sim 10-100 \rm cm^2 g^{-1}$. As a result, the merger-nova emission could be weakened and shifted toward softer bands. However, significant energy injection from the BP wind may suppress the production of lanthanides. In the following, we exhibit the results for both $\kappa=0.2 \rm cm^2 g^{-1}$ (green solid lines) and $\kappa=10 \rm cm^2 g^{-1}$ (red lines).

In the calculations, we fix the typical duration of central engine during prompt emission phase as $t_{90}=1$s. Therefore, the case with a larger initial disk mass denotes a higher initial accretion rate. For high $M_{\rm d}$ (red solid lines), the ejecta can be accelerated to a higher speed since more BP wind energy is available to drive the ejecta (see Figure \ref{fig2}). It thus becomes transparent and then peaks in luminosity at an earlier time compared to the low $M_{\rm d}$ case (red dotted lines). As a result, the merger-nova with disk mass $M_{\rm d}=0.1 M_\sun$ is brighter than that with $M_{\rm d}=0.01 M_\sun$. Compared to the high-mass ejecta (red dashed lines), a low-mass ejecta (red solid lines) is easier to be accelerated, which leads to a smaller peak time, larger peak energy $\epsilon_{\gamma,\rm p}$ and a brighter merger-nova.

In Figure \ref{fig2}, we also show the case with fall-back accretion (blue solid lines). We adopt $t_0=1000$s, $t_{\rm p}=1500$s and $\dot{M}_{\rm p}=6\times 10^{-5} M_{\sun}~s^{-1}$ in the calculations. The BP energy due to fall-back accretion would further heat up and accelerate the ejecta at late times, which lead to an even brighter and broader merger-nova.

\begin{figure}[htp]
\center
\includegraphics[width=8cm,angle=0]{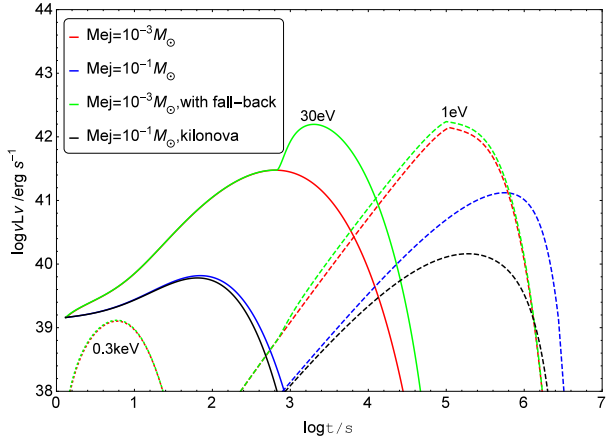}
\caption{The lightcurves of BP-powered merger-nova at different observational frequencies (dotted lines: 0.3keV, solid lines: 30eV, dashed lines: 1eV) for different ejecta mass (red lines: $M_{\rm ej}=10^{-3}M_{\odot}$ and $\alpha_{\rm m}=1$, blue lines: $M_{\rm ej}=0.1M_{\odot}$ and $\alpha_{\rm m}=1$). The model predictions with fall-back accretion are shown with green lines ($M_{\rm ej}=10^{-3}M_{\odot}$ is used). For comparison, we also show the predictions of kilonova ($\alpha_{\rm m}=0$, black lines). Other parameters are $M_{\rm d}=0.1M_{\odot}$ and $\kappa=10\ \rm cm^2 g^{-1}$.}

\label{fig3}
\end{figure}

Figure 3 presents the lightcurves of the BP-powered merger-nova at different frequencies (dotted lines: 0.3keV, solid lines: 30eV, dashed lines: 1eV).
The peak luminosity in the optical band ($\sim$1eV) can reach $\sim 10^{42}\rm erg~s^{-1}$. One may expect a brightening at $t>10^3$s for ultraviolet ($\sim$30eV) lightcurve of a low mass ejecta if there is fall-back accretion. However, the X-ray ($\sim$0.3eV) emission is generally too weak for the current detectors.

\begin{figure}[htp]
\center
\includegraphics[width=8cm,angle=0]{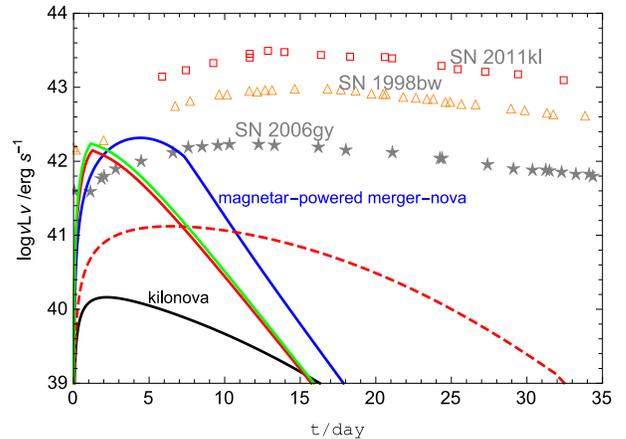}
\caption{A comparison of the optical ($\sim$1eV) lightcurves of the BP-powered merger-nova with supernovae (SN 1998bw and SN 2006gy), superluminous supernovae (SN 2011kl), kilonova ($\alpha_{\rm m}=0$, black line) and magnetar-powered merger-nova (blue line). The red solid line and red dashed line represent $M_{\rm ej}=10^{-3}M_\sun$ and $0.1 M_\sun$, respectively. For both cases, we use $\alpha_{\rm m}=1$. The green line corresponds to the BP-powered merger-nova emission with fall-back accretion ($M_{\rm ej}=10^{-1}M_\sun$ is taken). We adopt the parameters $\kappa=10 \rm cm^{2} g^{-1}$, $\beta=0.15$. For BH, we use the parameters $M_{\rm d}=0.1M_{\odot}$ and $\dot{M}_{\rm p}=6 \times 10^{-6} \rm M_{\odot} s^{-1}$. For the magnetar case, the parameters are $M_{\rm ej}=10^{-2}M_\sun$, $B_{15}=5$, and $P_{\rm i,-3}=5$.}

\label{fig4}
\end{figure}

Finally, in Figure \ref{fig4}, we compare the optical lightcurve of the BP-powered merger-nova (red lines: without fall-back accretion, green line: with fall-back accretion) with those of supernovae (SN 1998bw and SN 2006gy), superluminous supernova (SN 2011kl), kilonova ($\alpha_{\rm m}=0$, black line) and magnetar-powered merger-nova (blue line). The injection energy of magnetar model is dipole radiation (Zhang \& M\'{e}sz\'{a}ros 2001)
\begin{eqnarray}
L_{\rm sd}=L_{\rm sd,0}\left(1+{t\over t_{\rm sd}}\right)^{-2}
\end{eqnarray}
with $L_{\rm sd,0}=10^{49}~R_{s,6}^6B_{15}^{2}P_{i,-3}^{-4}\rm ~erg~s^{-1}$ and $t_{\rm sd}=10^{3}~R_{s,6}^{-6}B_{15}^{-2}P_{i,-3}^{2}\rm s$. The parameters adopted here are $B_{15}=5$ and $P_{i,-3}=5$, $R_{s,6}=1.2$, which gives $L_{\rm sd,0}=1.2\times10^{48}\rm ~erg~s^{-1}$ and $t_{\rm sd}=1.1\times10^{3}$s.

Inspecting Figure \ref{fig4}, we find that the injected energy from both BH and magnetar would enhance the luminosity of the merger-nova. For a new-born BH with strong BP-wind (high $\alpha_{\rm m}$), the merger-nova can be two orders of magnitude brighter than a kilonova. Therefore, the detection of a GW event, in which the post-merger product is a BH while the merger-nova is bright, would suggest that the BP mechanism is indeed at play.
Another feature of this model is that it may produce a bright merger-nova for an on-axis weak SGRB (corresponding to the low spin case).


\section{Conclusions and Discussions}
The mergers of NS-NS and NS-BH binaries are candidate sources of GWs. Much attention has been paid on the near-isotropic EM emissions from such events. A faction of the merger products could be a stellar mass BH. In this Letter, we found that the BP driven wind from this new-born BH-disk can produce a bight merger-nova peaking in the optical band. The peak luminosity can reach $\sim 10^{42} \rm erg\ s^{-1}$ in the hour-day timescale.

The ultraviolet-optical-infrared (UVOIR) transient (AT2017gfo) of GW170817 can be successfully interpreted with the theory of r-process ``kilonova''. The comprehensive observations on AT2017gfo reveals the details of light curves and spectral evolution (Kasen et al. 2017). A single component kilonova is found difficult to account for the early and late data as well as the evolution of the spectra. Villar et al. (2017) collected the available and homogenized optical data and found that the data can be well modeled with a three-component kilonova model. Yu \& Dai (2017), on the other hand, proposed a hybrid energy sources including radioactivity and NS spin-down to explain the complex optical emission of AT2017gfo. However, if the merger product is a BH or a short-lived NS followed by prompt collapse to a BH, a hybrid model with a r-process and energy injection from BP-wind might be relevant.

The BP-powered merger-nova can be as bright as the magnetar-powered merger-nova. The late-time fall-back accretion might further enhance the luminosity of the merger-nova. The BZ power has strong dependence on the BH spin $a_\bullet$, while the BP power is insensitive to $a_\bullet$. One would expect an on-axis low-luminosity SGRB with a bright merger-nova if $a_\bullet$ is quite small. Combining the observations of GWs, SGRBs and merger-nova emissions, we can have a better understanding of the origin and merger product of the GW events. In particular, if the GW data show that the post-merger product is a BH while the merger-nova is bright, one would suggest that the BP mechanism is indeed at play. Such an event would server as a robust test of our model.

Besides the merger-nova emissions, the BP driven wind can also produce multi-band afterglow radiation in the ``free zone'' (the direction where the SGRB is missed, but X-rays may escape, see Sun et al. 2017). Such emission is is potentially detectable by future wide-field X-ray detectors, such as Einstein Probe (EP) (Yuan et al. 2016). The coexistence of a BZ jet and a BP wind may develop a structured jet, which may give rise to a low-luminosity SGRB for a slightly off-axis observer, which may be the case of GRB 170817A (Xiao et al. 2017; Zhang et al. 2017).

Significant energy injection due to mass loss in a merger system was studied by Song \& Liu (2017), who pointed out that the resulting kilonova event may be as bright as a supernova. Due to the uncertainties of the mass loading in the outflow, we ignored energy injection from wind itself in this work. 

Another consideration is that the merger-nova photons would be scattered into higher frequency via inverse compton (IC) in the cocoon shocked region (Ai \& Gao 2017). In the context of BH central enine, it is worth studying the IC scattered emission from BP-powered merger-nova photons.

\section*{Acknowledgements}

We thank Yunwei Yu for helpful discussions. The Numerical calculations were performed by using a high performance computing cluster (Hyperion) of HUST. This work is supported by the National Basic Research Program ('973' Program) of China (grants 2014CB845800), the National Natural Science Foundation of China under grants U1431124 and 11773010. HG acknowledges support by the National Natural Science Foundation of China under grants 11722324, 11603003, 11690024 and 11633001.


\begin{thebibliography}{24}
\bibitem{}Abbott, B. P., Abbott, R., Abbott, T. D., et al. 2016, Physical Review Letters,116, 241103
\bibitem{}Abbott, B. P., Abbott, R., Abbott, T. D., et al. 2016, Physical Review Letters,116, 061102
\bibitem{}Abbott, B. P., Abbott, R., Abbott, T. D., et al. 2017a, Physical Review Letters,119, 161101
\bibitem{}Abbott, B. P., Abbott, R., Abbott, T. D., et al. 2017b, ApJL, 848, L13
\bibitem{}Ai, S. K., \& Gao, H. 2017, submitted
\bibitem{}Bardeen, J.~M., Press,W.~H., \& Teukolsky, S.~A.\ 1972, \apj, 178, 347
\bibitem{}Blandford, R. D., \& Payne, D. G., 1982, MNRAS, 199, 883
\bibitem{}Blandford, R. D., \& Znajek, R. L., 1977, MNRAS, 179, 433
\bibitem{}Bucciantini, N., Metzger, B. D., Thompson, T. A., Quataert, E. 2012, MNRAS, 419, 1537
\bibitem{}Burrows D.~N., Romano P., Falcone A., et al. 2005, Science, 309, 1833
\bibitem{}Burrows, D. N., Grupe, D., Capalbi, M., et al. 2006, ApJ, 653, 468
\bibitem{}Chen, W., Xie, W., Lei, W. H., Zou, Y. C., L\"{u}, H. J., Liang, E. W., Gao, He., \& Wang, D. X. 2017, ApJ, 849, 119
\bibitem{}Dai, Z.~G., \& Liu R.-Y.\ 2012, \apj, 759, 58
\bibitem{}Fan, X. L., Messenger, C., \& Heng, I. S. 2017, arXiv:1706.05639
\bibitem{}Fong W., Berger E., Servillat M., Anglada G., et al. 2013, ApJ, 769, 56
\bibitem{}Fox, D. B., Frail, D. A., Price, P. A., Kulkarni, S. R., Berger, E., et al. 2005, Nature, 437, 845
\bibitem{}Gao, H., Ding, X., Wu, X. F., Zhang, B., \& Dai, Z. G. 2013, ApJ, 771, 86
\bibitem{}Gao, H., Ding, X., Wu, X.-F., Dai, Z.-G., \& Zhang, B. 2015, ApJ, 807, 163
\bibitem{}Gao, H., Zhang, B., \& L\"{u}, H.~J. 2016, Phys. Rev. D 93, 044065
\bibitem{}Gao, H., Zhang, B., L\"u, H.-J., \& Li, Y. 2017, ApJ, 837, 50
\bibitem{}Gehrels, N. et al. 2005, Nature, 437, 851
\bibitem{}Jin, Z.-P., Li, X., Cano, Z., et al. 2015, ApJL, 811, L22
\bibitem{}Jin, Z.-P., Hotokezaka, K., Li, X., et al. 2016, Nature Communications, 7, 12898
\bibitem{}Kasen, D., Bildsten, L. 2010, ApJ, 717, 245
\bibitem{}Kasen, D., Badnell, N. R., \& Barnes, J. 2013, ApJ, 774, 25
\bibitem{}Kasen, D., Metzger, B., Barnes, J., Quataert, E., \& Ramirez-Ruiz, E. 2017, arXiv:1710.05463
\bibitem{}Korobkin, O., Rosswog, S., Arcones, A., Winteler, C. 2012, MNRAS, 426, 1940
\bibitem{}Kumar, P., Narayan, R.,\& Johnson, J.~L.\ 2008, \mnras, 388, 1729
\bibitem{}Lei, W. H., Zhang, B. \& Liang, E. W. 2013, ApJ, 756, 125
\bibitem{}Lei, W.~H., Zhang, B., Wu, X.~F., \& Liang, E.~W. 2017, ApJ, 849, 47
\bibitem{}Li, L.-X., \& Paczy\'nski, B. 1998, ApJ, 507, L59
\bibitem{}Li Y., Wang D. X., \& Gan Z.M., 2008, A \& A, 482, 1
\bibitem{}Liu, T., Gu, W. M., {\&} Zhang, B. 2017, New Astronomy Review, 79, 1
\bibitem{}Livio, M., Ogilvie, G. I., \& Pringle, J. E. 1999, ApJ, 512, 100
\bibitem{}L\"{u}, H.~J., Zhang, B., Lei, W.~H., Li, Y., Lasky, P.~D., 2015, ApJ, 805, 89
\bibitem{}MacFadyen, A.~I., Woosley, S.~E., \& Heger, A.\ 2001, \apj, 550, 410
\bibitem{}Meier, D. L. 2001, ApJ, 548, L9
\bibitem{}Metzger B. D., Martinez-Pinedo G., Darbha S., Quataert E., Arcones A., Kasen D., Thomas R., Nugent P., Panov I. V., Zinner N. T., 2010, MNRAS, 406, 2650
\bibitem{}Metzger, B.~D., \& Piro, A.~L.\ 2014, \mnras, 439, 3916
\bibitem{}Novikov, I. D. \& Thorne, K. S., 1973, in Black Holes, ed. Dewitt C (Gordon and Breach, New York) p.345
\bibitem{}Song, C. Y., \& Liu, T. 2017, arXiv:1710.00142
\bibitem{}Sun, H., Zhang, B., \& Gao, H. 2017, ApJ, 835, 7
\bibitem{}Villar. V. A., Guillochon, J., Berger, E. et al. 2017, arXiv:1710.11576
\bibitem{}Virgili, F., Zhang, B., O'Brien, P., Troja, E. 2011, ApJ, 727, 109
\bibitem{}Wu, X.~F., Hou, S.~J., \& Lei, W.~H.\ 2013, \apj, 767, L36
\bibitem{}Xiao, D., Liu, L.-D., Dai, Z.-G., Wu, X.-F. 2017, arXiv:1710.00275
\bibitem{}Yu, Y.-W., Zhang, B., \& Gao, H.\ 2013, APJL, 776, L40
\bibitem{}Yu, Y.-W., \& Dai, Z.-G. 2017, arXiv:1711.01898
\bibitem{}Yuan, W., Amati, L., Cannizzo, J. K., et al. 2016, SSRv, 202, 235
\bibitem{}Zhang, B. \& M\'{e}sz\'{a}ros, P. 2001, ApJ, 552, L35
\bibitem{}Zhang, B., Yan, H. 2011, ApJ, 726, 90
\bibitem{}Zhang, B. 2013, ApJL, 763, L22
\bibitem{}Zhang, B. B., Zhang, B, Sun, H., Lei, W. H., Gao, H., et al. 2017, arXiv:1710.05851




\end{thebibliography}
\end{document}